# Classroom Technology Deployment Matrix: A Planning, Monitoring, Evaluating and Reporting Tool.

Philip Heslop[1], Ahmed Kharrufa[1], Madeline Balaam[2], David Leat[1]

[1]Newcastle University, Newcastle upon Tyne NE1 7RU, UK

[2]KTH Royal Institute of Technology, SE-100 44 Stockholm, Sweden

**Abstract.** We present the Classroom Technology Deployment Matrix (CTDM), a tool for high-level Planning, Monitoring, Evaluating and Reporting of classroom deployments of educational technologies, enabling researchers, teachers and schools to work together for successful deployments. The tool is derived from a review of literature on technology adaptation (at the individual, process and organisation level) – concluding that Normalization Process Theory, which seeks to explain the social processes that lead to the routine embedding of innovative technology in an existing system, would a suitable foundation for developing this matrix. This can be leveraged in the specific context of the classroom – specifically including the Normal Desired State of teachers. We explore this classroom context, and the developed CTDM, through looking at two separate deployments (different schools and teachers) of the same technology (Collocated Collaborative Writing), observing how lessons learned from the first changed our approach to the second. The descriptive and analytical value of the tool is then demonstrated through mapping these observation to the matrix and can be applied to future deployments.

## 1    Introduction

The challenge of taking educational technology out of the lab and deploying it "in the wild" in classrooms is not only technological (will it work?) or even pedagogical (will it assist teaching?), but also cultural and practical (will the school and teachers accept it?) [6]. This work focuses on the high level deployment process, a rather than evaluating the technology itself. Technology adoption in the classroom is often characterised as the teachers' responsibility: "teachers' mind-sets must change to include the idea that 'teaching is not effective without the appropriate use of technology to facilitate student learning.'" [12]. Teachers disposition towards technology is reported as being affected by many factors, including their "existing attitudes and beliefs toward technology, as well as their current levels of knowledge and skills." [13,34], however this emphasis places the onus on teachers' attitudes to technology and detaches the adoption of technology from teachers practice (i.e. teaching). Part of the challenge of a successful adoption of technology is complimenting teacher's practice in such a way that the teacher sees the benefits and does not feel undermined or usurped. As such, a good starting point is gaining an understanding of the teachers' current views [4].

'Normal Desirable State' (NDS) [7] is a model of the teachers' current perspective of the desired state of pupil activity in the classroom. It is how a teacher understands progress (pupils learning and development, or creation of products or coverage of work), and provides criteria for teachers to self-evaluate a lesson and make decisions about their actions based on immediate feedback. NDS varies across different teachers, across different activities and across different stages of a lesson. The NDS model however does not make clear the role of technology – as either an enabler or inhibitor.

However, there is a balance to be struck here. Focusing too much on the performance of the technology may undermine teachers' desired state or school norms, i.e. when asking them to act outside their NDS. Recognising and capturing the desired state, and the tensions created by moving teachers beyond it, could increase the likelihood of a successful deployment. This goes beyond the concepts of 'training' or 'changing teachers mind-sets' [12,40], which offer partial solutions but reinforce a perceived power dynamic where the researchers/technology are 'right' and the school and teachers' existing practices are deficient in some way.

Therefore a full understanding of the NDS of the school and teachers and incorporation of these into the study design is likely to increase the chances of success and minimises tensions. We support this hypothesis through analysing and reflecting upon the outcomes of two large-scale classroom deployments of the same technology, in two schools that represent different contexts [17,18,24]. The concept of NDS of the teachers and the school was not explicitly considered in the first, but was taken into account for the second. We define success through the level of integration of the technology into the teacher's practice and lessons beyond the study. We consider several technology adoption models that acknowledge that the technology needs to integrate and compliment teachers practice and beliefs.

In this paper, we contribute a Classroom Technology Deployment Matrix (CTDM), integrating NDS and technology adaptation, for longitudinal, classroom-wide, teacher facilitated educational technology deployments. CTDM is a tool that facilitates high-level Planning, Monitoring, Evaluating and Reporting for these deployments. It is aimed at specific actors in the process, researchers, teachers and schools, but also takes into account the input of students. It is a tool designed to enable researchers, teachers and schools to work together for successful deployments.

## 2    Related Work

Research on classroom technology concentrates on the effect of the technology without discussing deployment. Deployments on a full classroom scale similar to those explored in this work also tend to be less teacher-led interventions, such as the SOLE (Self Organised Learning Environment) project [11], and Think Active [15], or explore new technology-enabled modalities of learning such as embodied interaction [26,38]. Where technology is more teacher focused, there tends to be a focus on technology to manage or orchestrate sessions in a classroom [1,20,29,30], rather than covering the full span of the work (i.e. incorporating teachers NDS, planning, deployment, sessions and analysis).

While there may well be a need for upskilling or training teachers [40] as part of a wider set of technology adoption in the classroom, the problem of deployment needs to take a wider view – that of empowering teachers and schools to influence the design and implementation details of deployments. The SAMR and TPACK models (often used in conjunction) [16,19,39,44] go some way to get beyond the "training" metaphor and incorporate teachers own pedagogical approaches. However, these approaches focus on assisting teachers with existing technology integration [19], rather than researching the role of technology in pedagogy, including deployment of novel learning technology, which may challenge the pedagogical philosophy of the researchers, teachers (or school). They do not incorporate the teachers' (or school's) input into the design or planning phase of a deployment, and do not address change over time. The models can, however, be incorporated into a larger Planning, Monitoring, Evaluating and Reporting process, and provide a proven and robust mechanism for this purpose.

### 2.1    Technology Adoption Models

While the area of "technology adoption" is a vibrant research area in itself, it is important to consider the characteristics of a classroom deployment, and how it differs from other, more common situations (e.g. medical settings [2]). A deployment is a shared task between the researchers, school, teachers and students, and goes beyond the concept of training or intervention (which implies a power dynamic with researchers in control) [40], as fundamental pedagogical and organisational concepts may well be challenged which can meet resistance regardless of the technology.

A common methodology in technology adaptation is the concept of domestication [3], i.e. technology over time becomes tamed, integrated and habitual. There is a strong link between this "domestication" and the concept of "home". While the idea of having technology becoming as familiar as everyday appliances is appealing, it is beyond the time-limited scope of most educational technology deployments, and the classroom is not as comforting or familiar as home. The technology acceptance model [9,46], on the other hand, is focused on the perceived usefulness, ease of use, and user acceptance of technology, and, while having some use in educational technology applications [14], is focused on end users (i.e. students) rather than those in supervisory or supplementary roles (i.e. teachers). The concept of Technological Pedagogical Content Knowledge [36] is aimed more at teachers, but focusing on "non-classroom tasks" (i.e. teacher-run pedagogical systems, e.g. online course design, educational websites, media creation etc.), rather than the dynamic classroom role they would take in most deployments.

While these models may provide partial solutions, they do not encompass all the requirements for a deployment. We require a model that takes a higher-level approach that acknowledges the end users of technology but also considers how the overall system may be affected – i.e. how this corresponds to the teachers NDS. Normalization Process Theory (NPT) [31] offers a high-level model that attempts to "understand the things that individuals and groups do to operationalize new or modified modes of practice as they interact with dynamic elements of their environments". NPT is concerned with agents (i.e. people) in a system, and how they adapt to changes in a system through the following mechanisms:

a. **Coherence** - how a practice is made possible by peoples' ideas of and investment into (individual and social) its meaning;
b. **Cognitive Participation** - how people act to initiate and be enrolled into delivering a group of practices;
c. **Collective Action** - how people enact those practices (and how is it understood?);
d. **Reflexive Monitoring** - how people act to appraise the consequences of their contributions.

These mechanisms are loosely aligned with the phases of a deployment [28], i.e. Coherence is aligned with the status quo, Cognitive Participation with planning a deployment, Collective Action with implementation and Reflexive Monitoring with reflection. However, the mechanisms can and do occur throughout the process.

NPT has been widely used to integrate change (and in particular technology) into various medical organisations [32,33], and works because it acknowledges the existing expertise and processes of "agents" (people) and the overall aims and goals of the organisation as well as the dynamic nature of a deployment. We apply a similar process to education technology integration into the classroom.

In this paper, we investigate two related studies (using the same technology) through this lens to capture the disposition of the various parties and demonstrate how this lens helps in identifying shortfalls in the first study and brings about what contributed to the success of the second. In doing so, we use our analysis to construct a combined NDS/NPT classroom matrix, a process model for Planning, Monitoring, Evaluating and Reporting on technology deployments in the classroom.

## 3    Background

Collocated Collaborative Writing (CCW) [17,18] is a Digital Tabletop application for the collaborative learning of a persuasive writing task. It has been previously evaluated as a learning tool [17,18,22]. It is designed to exploit the benefits of socio-cognitive learning [27,48] and distributed cognition [10,24,37,41], exploiting scaffolding and fading [50,51] by allowing learners to externalise and communicate their thinking visuospatially. The benefits of shared visuospatial representations include reduced cognitive load through externalisation, a deeper understanding of the problem through re-representation and a means of distributing thoughts and ideas between collaborators [25,37]. In order to evaluate CCW, the technology was taken out of the lab and deployed in classrooms, requiring the transport and setup of eight digital tables, which had to be dismantled between sessions due to the classrooms being used for other activities. To generate ambiguous topics to write about persuasively, Digital Mysteries [21,24] - a digital tabletop application - was used, with materials (i.e. mysteries) created by the teachers.

### 3.1 Studies

**Study One (School One)**
Participants were students of mixed ability, year 8 (aged 13-14), studying English, Geography and History, across two classes (Table 1). Five teachers were involved, two geography, two history and one English. Due to scheduling issues, a single classroom was unavailable for every session, and some sessions were conducted in a less suitable ICT room.

**Table 1.** Study One

| Classes | Students Per Class | Teachers | Subjects | Tables | Sessions |
|---------|--------------------|----------|----------|--------|----------|
| 2 | ~30 | 5 | 3 | ~ 8 | Class A: 3 Class B: 2 |

Before deployment, several pre-study activities took place. Group-work lessons were observed and videoed in order to establish the kind of group based activities the students would normally be involved in and give insight into school culture. Teachers were invited to try out the technology at the university before the classroom deployment. Students were asked to complete a Pupil View Template (PVT) [49] exercise, to ascertain their disposition to: Learning in the Classroom, Working in Groups, Working on a Problem and Working in Groups around a digital tabletop.

Each class completed the study across 2 or 3 subjects class B did History and Geography, while class A did History, Geography, and English. The students' usual teachers facilitated, and groups were organised by the teachers into mixed ability groups. Students first completed a Digital Mysteries [24] exercise followed by a Collaborative Writing [18] exercise using the same evidence. Sessions were recorded on two video cameras (classroom level and group level). Post-study activities included: semi-structured interviews with four groups of students and the teachers; a semi-structured group interview with most of the teachers, and students also completed an exercise to ascertain their disposition to the study, mirroring the earlier PVT task.

**Study Two (School Two)**

We worked with one English teacher and a single mixed ability class (year 8, 13-14 year olds) (Table 2).

**Table 2.** Study Two

| Classes | Students Per Class | Teachers | Subjects | Tables | Sessions |
|---------|-------------------|----------|----------|--------|----------|
| 1 | ~30 | 1 | 1 | ~ 8 | 4 |

Before the study, the teacher and researchers conducted planning meetings, where the capabilities of the technology, the teacher's aspirations for the study and initial lesson plans designed to incorporate the technology into the teaching agenda were discussed. Similar meetings occurred throughout the study, and the teacher provided written reflections and updated lesson plans for each session. This ongoing involvement of the teacher happened organically and did not follow a planned co-design program (e.g. [8,42,43]. Sessions took place in a single secondary school classroom in four sessions over 6 weeks. The classroom was equipped with 8 smart tables, with groups of 3-4 students at each – 30 students in total. Groups were mixed ability and consistent across the study, selected by the teacher. The teacher facilitated each lesson, within the existing timetable. 2-3 Researchers were also present, and sessions were filmed. Before each writing session, the students completed a collaborative exercise, either a Digital Mystery [23] (first 3 sessions) or a classroom debate (for the final session).

### 3.2 Methodology

**Table 3.** Data Collected from Studies

| | School 1 | School 2 | Purpose |
|---|----------|----------|---------|
| Teacher Technology Sessions (Researcher Notes) | X | X | To familiarise the teacher(s) with the digital tabletop technology |
| Teacher Lesson Planning Sessions (Conducted throughout study) | | X | To integrate the technology into the formal lesson plans (i.e. assigning specific learning goals to activities) |
| Pre-Study Group-work Video | X | | To show existing group-work practice |
| Pupil View Templates | X | | To ascertain disposition of students towards group-work, technology etc. |
| Session Videos | X | X | To show main study activity |
| Post Study Teacher Interviews | X | X | To ascertain teacher(s) view of the study |
| Post Study Student Interviews | X | X | To ascertain students view of the study |

Due to the realities of conducting studies "in the wild" (time, teacher availability and space pressures), and the refinement process that occurred between the studies, the kinds of data collected were not identical. (Table 3)**.**

We conducted an *inductive* thematic analysis [5] on data collected in the first study. The analysis followed a standard thematic analysis process with two researchers, R1 & R2:

1. R1 Transcribed & Contextualised data from all sources (i.e. video, audio and documents – table 3), and this was read by both Researchers independently.
2. Codes were generated independently by each researcher and then consolidated,
3. Data was classified into candidate themes,
4. Themes were finalised and defined.

This resulted in the following themes: 1) responsibility and expectation, 2) culture, 3) technology for teachers, and 4) resources and priority. Two Researchers used these themes to conduct a *deductive* thematic analysis on the second study.

These evaluations showed that the second deployment was more successful (i.e. the level of integration of the deployment technology into the teacher's practice outside the study).

We then surveyed common Technology Adaptation models, choosing Normalization Process Theory (NPT) [31], and correlated our themes with the generalised model to produce a refined model aimed at large-scale classroom deployments (Figure 1).

To help validate the model, we conducted semi-structured interviews with two experienced HCI researchers working on classroom deployments of technology (that had significant teacher facilitation and input [15,47]).

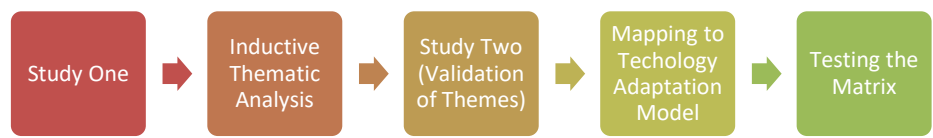

**Fig. 1.** Research Process

## 4 Results and Analysis

We present the themes generated from the initial thematic analysis (1) responsibility and expectation, 2) culture, 3) technology for teachers, and 4) resources and priority. We explain their provenance from the data and correlate this with the analysis from the second study. We identify the differences that lead to a more successful outcome (the level of integration of the deployment technology into the teacher's practice and lessons outside the study), and to indicate where these differences are under the control of the research team.

## 1-Responsibility and Expectation

In study one, there was a discrepancy between what the researchers, teachers and students assumptions about what aspects of the study were under whose jurisdiction. Generally, teachers assumed that the technology would do more than it was capable of – including regulating behaviour. Conversely, students were expecting the teachers to have more input in orchestrating the lesson. The perceptions of the three groups are captured in Table 4 (where T=Teachers, R=Researchers/Technology, S=School), showing what each party was expecting from the others (and where these expectations were mis-matched).

**Table 4.** School One: Stakeholders Perceptions of who is Responsible? – R = Research Team, T = Teachers, S = School

|  | Researchers | Teachers | School | Students |
|---|---|---|---|---|
| Who is responsible for assuring the room and schedule are correct? | T/S | R | R | T |
| Who is responsible for the setup and management of tables? | R | R | R | T |
| Who is responsible for integrating technology into the learning plan? | T | R | T | T |
| Who is responsible for the learning materials? | T | R/T | T | T |
| Who is responsible for differentiation and scaffolding? | T | R | T | T |
| Who is responsible for progression of the task? | T | R | T | T |
| Who is responsible for regulating behaviour in the classroom? | T | R/T | T | T |
| Who is responsible for Assessment? | T | R | T | T |

This is impactful on the NDS, as teachers are unsure of pupils' progress, how to evaluate their own input and exactly what actions they could perform to change the current state. One cause of this discrepancy is that the research team assumed too much about the teachers' knowledge about the limitations of the technology. The hands-on sessions with the teachers focussed on the advantages of the technology, and did not help define the role of the teachers in this "new" context (e.g. to regulate behaviour, provide orchestration and more importantly to continue to assess progress and provide feedback regardless of the technology). In particular, the pupil view template exercises illustrated the students' attitude towards group work, technology and the responsibility of the teachers during these lessons. Students often reflected on teachers being unfamiliar with the technology and not being able to recognize what was correct behavior. There was also an indication that group work allowed teachers to be 'lazy' (i.e. by not being at the front of class and lecturing) which was reflected by teachers in interviews indicating that group work was a way for students to 'hide'.

In study two, the relationship with the teacher was more thorough. Lesson planning meetings were held regularly, and scenarios (both potential and experienced from previous sessions) were worked through with the teacher to help define the boundaries between what the technology can do and what the teacher needs to do (and where they should work together). The pupils completed short feedback exercises that indicated they were supportive of the teacher (who spent more time explaining at the front of the class) and the group work, reserving their criticism for the technology itself.

## 2-Culture

While ostensibly following the same curriculum, the two institutions had differing teaching philosophies. School one focused on imparting knowledge and assessment. Although developing thinking skills were part of the schools official remit, the teaching of them was given a lower priority. In fact, several teachers were skeptical about the students' ability to comprehend thinking skills or develop transferrable skills - *"skills are not transferable between subjects"* and *"if a student learns how to write an argument in history, they would have to be also taught that skill in geography"*. This clashes with the design principles of the technology (socio-cognitive learning), creating tension that affects the disposition of the teachers towards the technology. In lessons, students notice this discomfort and behave disruptively without the teacher being confident enough to chastise them. i.e. the NDS of the teachers was challenged, not just at the technology imposition level but at the pedagogical level – teachers were not equipped to assess progress, evaluate their teaching or decide on actions in-line with the pedagogical grounding of the technology.

The second school had a different education culture and approach to teaching thinking skills. It was seen as a useful and worthwhile endeavor, although there were still doubts about assessment. The teacher involved in this study considered skills such as writing convincing arguments as useful across multiple subjects. This led to the teacher having a more positive regard for the technology as the underlying principles more closely matched her beliefs. This was also helped by the teacher having ongoing input into how the technology was used so that it matched with her teaching goals.

One of the main pedagogical tensions that became apparent in the two studies was the balance between *imparting knowledge* (acquisition metaphor, didactic, convergent & easy to assess) and *competence development (*participation metaphor, problem solving, divergent and difficult to assess [45]).

## 3-Technology for Teachers

One way technology could assist NDS is to provide orchestration and monitoring tools for teachers. In the first study, the researchers focused on providing a learning tool for the students. Teachers were involved in the planning process but there was little incentive to "own" the technology - there were only two "familiarisation" sessions at the University rather than at the school and the design was seen as final and immutable. Some were encouraged to volunteer by the head teacher on the basis that they would find it interesting and that it would help with their career development to "be involved in research". In short, the teachers worked round the technology, trying to minimise impact on their NDS and not taking advantage for their own agenda. They only had 1 or 2 sessions using the technology and so were unable to build up a rapport (a trusting relationship) with it – they could not see progress or adapt their plans to the changing dynamic the technology provided. This wariness emerged in the classroom, with statements to pupils like *"I'm not an expert on using this"* or *"You are in the same boat as me with the technology"*. The students, looking to the teachers for leadership around this new experience, were met with doubts and even cynicism. This came out in their pupil view template exercises.

In study 2, researchers developed a relationship with a single teacher across several sessions in school and consulted initially on how they thought the technology would benefit them. The design of the software was tweaked based on the teachers feedback, with the main improvement being a more concentrated focus on the planning element leaving the "text-generation" as an individual (and assessable) exercise – i.e. the design was seem as changeable by the teacher to fit with their teaching goals and NDS. Lesson plans incorporated the affordances of the technology (e.g. supporting group work, and providing structure to a process) as well as the lesson topic. The teacher provided feedback after each session, and adapted plans to how the students were responding to the technology. The teacher became more comfortable and required less class-wide interventions, instead triaging group and individual issues, confident that the class as a whole was progressing according to her plan.

## 4-Resources and Priority

Schools have limited resources, with space and time at a premium, and prioritise what they feel is most useful to pupil progress (i.e. what is best for NDS). The study required that there was significant preparation time (~ 1 hour), and a similar dismantling process (layout impacted the lessons [35]). During the first study, the deployment was moved to whatever space was available, which in some cases was in retrospect unsuitable, i.e. in a computer lab with existing equipment and little seating space (Figure 2). There were also issues with scheduling with several sessions cancelled for higher priority activities, e.g. internal exams. With little redundancy in the programme, this meant the study was shorter than planned and there were large gaps between sessions. These real-world scenarios are often overlooked when planning a deployment.

For the second study, a longer time-period allowed for potential scheduling issues, as well as a longer planning stage with the teacher. It was agreed that the same classroom would be used for every session. This led to a smoother study, although the classroom was in use between sessions so the set-up and dismantling process was still required for every session. Without the experience of the first study, these considerations would have been missed and similar scheduling and space issues would have occurred.

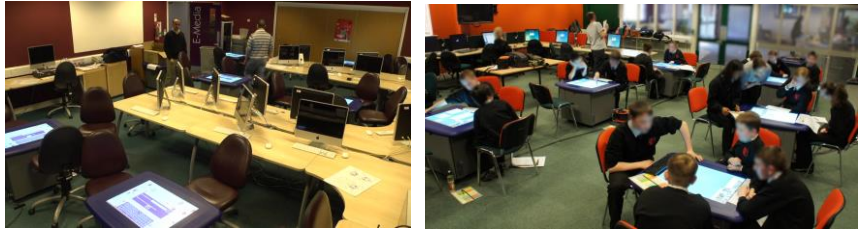

**Fig. 2.** Deployment in a computing lab vs large classroom

## 5 Classroom Technology Deployment Matrix

Using the following shorthand for each of the actors involved (T=Teachers, S = School, R = Researchers, St = Students). We mapped the emerging themes from the studies to the mechanisms suggested in the NPT model to produce the following Classroom Technology Deployment Matrix (CTDM) (Table 5).

The generalised NPT model could be mapped onto the processes involved in a classroom deployment as follows:

a. Coherence – what is the researchers, teachers, school and students **initial disposition** towards the proposed deployment? Is it aligned with, or does it challenge their NDS?

b. Cognitive Participation – how involved (and invested) are the researchers, teachers, school and students in the **planning and design** of the deployment? Do they have control over how it may impact their NDS?

c. Collective Action – how do researchers, teachers and students engage with the **ongoing** deployment? Does their disposition change?

d. Reflexive Monitoring – how do researchers, teachers and students reflect on the deployment, **during and after**?

The matrix (Table 5) provides a *high level* framework for a deployment, it is agnostic as to the specific mechanisms by which each of the elements is addressed. e.g., the SAMR/TPACK models [19,39] can be adapted to form part of the process, i.e. fulfilling elements in row 3, Technology for the Teacher.

**Table 5.** Classroom Technology Deployment Matrix – T = Teacher(s), S = School, St = Students, R = Researchers

| | **(a) Coherence** | **(b) Cognitive Participation** | **(c) Collective Action** | **(d) Reflexive Modelling** |
|---|---|---|---|---|
| **1. Responsibility & Expectation** (Who should be doing what?) | In the classroom, what are: **T:** Perception of **current** role **S:** Expectations of Teacher **St:** Disposition to learning | What involvement do **T, S, St** and **R** have in the **initial design** of the intervention? i.e. what design methodology is used? | What involvement do the **T, S, St** and **R** have in the **running and ongoing design** of the intervention? i.e. what is the scope for **change**, and who can instigate it? | What **changes** in understanding occur about **T, S,** and **R** Responsibilities? i.e. have roles and activities adapted for **the Intervention**? |
| **2. Culture** (What are the underlying philosophies and principles?) | How does the current classroom culture (theories of learning, methods) **correlate** across actors (**T, S & R** – i.e. **Intervention's** design principles?) | How can the current classroom culture (theories of learning, methods) of **T, S,** and **R in-fluence/change**? the deployment i.e. Is there scope? | How can the current classroom culture (theories of learning, methods) of **T, S,** and **R modify** underlying principles of **the Intervention**? | What **changes** may occur about the theoretical or philosophical disposition of **T, S,** and **R**? |
| **3. Technology for the Teacher** (How can the intervention improve Teachers practice?) | **R**: How much of **Intervention** is **specifically aimed** to fulfil: **T**: current NDS **S**: Expectation of T (e.g. indicating progress, capturing outcomes) | Can **T** and **S** influence the **initial design** of the **Intervention** in order to fulfil their NDS (e.g. indicating progress, capturing outcomes) | Can **R, T** and **S modify** the **Intervention** during deployment to fulfil their NDS (i.e. what mechanisms exist and how easy is it to change the **Intervention** during deployment)? | Aside from learning outcomes, did the **Intervention** fulfil the needs of **R, T** and **S**? If not, what design changes could be made? |
| **4. Resources and Priority** (How much space and time is available) | **S**: Does the **Intervention** correlate with the resources (classrooms, time, and importance) available/familiar to **T**? | Is there capacity for **R, T & S** to change the **initial** resources (classrooms, time, and importance) of the **Intervention**? | Is there capacity for **R, T & S to modify** resources (classrooms, time, and importance) during the **Intervention**? | **S**: What **changes** may occur to the resources (classrooms, time, and importance) because of the **Intervention**? |

## 5.1 Testing the Model

**Table 6.** Differences between studies

| | Study One | Study Two |
|---|---|---|
| **1a** | T: Facilitates Lessons and Creates Materials, S: expects T to do this in line with curriculum, St: Expects T to be authority in classroom (i.e. moderate behaviour). | |
| **1b** | T: presented with a finalised design developed by R. | **T: able to critique and improve initial design.** |
| **1c** | T: facilitated, but were unsure about technology role. T: able to suggest changes. Did not adjust plans to take advantage of technology. | **T: facilitated and able to suggest changes to improve facilitation. Incorporated technology into lesson plans.** |
| **1d** | T: remained unsure of role of technology in classroom. | **T: enthusiastic about exploiting technology in the future.** |
| **2a** | S: correlation with Intervention (participation metaphor). T: differing philosophies (acquisition metaphor). | **S, T & Intervention: well correlated.** |
| **2b** | T & S: little scope for philosophical change (despite non-correlation between T & S (2a). R: open to change | **T,S & R: already have good correlation (no motivation for change)** |
| **2c** | Intervention: (on cultural level) only changeable minimally due to time constraints. | - |
| **2d** | T, S & R not influenced to change their standpoints, despite low correlation. | **T, S & R already well correlated.** |
| **3a** | Intervention: developed before presenting to teachers. | **T had input to develop design and incorporate NDS.** |
| **3b** | T: No scope to change initial design. | **T influenced additions to design.** |
| **3c** | T: Had inputs that changed ongoing design. | - |
| **3d** | R: Lessons learned incorporated into future designs. | - |
| **4a** | S: initially had capacity for deployment. | - |
| **4b** | S: in full control of resources. | - |
| **4c** | S: made significant changes (room changes, cancellations etc.) | **S: Schedule agreed beforehand and unchanged.** |
| **4d** | Resources reset each session. No long-term changes. | - |

Using CTDM, the *differences* between studies can be shown in Table 6.

To test the model beyond these two studies, we conducted semi-structured interviews with two researchers who led other classroom deployments (a microbit project using computer science undergraduates to facilitate teaching microbit programming in several secondary schools [47], and the ThinkActive project incorporating activity trackers into lessons on fitness and health in primary schools [15]). The researchers are experienced HCI researchers in educational technology and were chosen as the projects in question were recent classroom deployments of technology which depended on a high level of teacher facilitation and input. We focused on their process and how the model corresponds with their experiences, including how they might utilize it in the future.

We endeavored to capture how these projects might fit within the themes generated from our study analysis (i.e. responsibility, culture, technology for teachers and resourses), and how the matrix might have been utilized in their planning and monitoring, and how it might go on to be used for evaluation and reporting. We asked: how the projects were instigated (i.e. academic, school or 3rd party) - specifically how, when and to what extent the school and teachers were involved; about the deployment planning process and how much influence schools and teachers had (and if plans changed); how learning materials were created (by who?) and delivered; what major tensions occured during the deployment; what has changed in the classroom following deployment; and how the matrix tool above could be used to evaluate and report these deployments, and plan, monitor, evaluate and report on future deployments.

Both projects were instigated by 3rd parties (i.e. not researchers or schools), without pre-engagement with schools to establish coherence (i.e. CTDM column a) - both interviewees suggested this would be beneficial in the future -"*it's like we didn't have time to consider this*" or "*recruitment [of schools] came after we had decided what to do*".

Both studies recruited schools after much of the deployment was planned, and researchers indicated that school and teacher involvement in planning stage was low (i.e. teacher involvement, CTDM column b), leading to tensions around responsibilities (row 1) and culture (row 2) - "*Teachers were 'volunteered' by the school and just went along with it*".

In the microbit project there were significant culture clashes: "*the school didn't tell us that 'trainee teachers' (the undergraduates were not trainee teachers but computing science researchers) were expected to wear suits and ties until the deployment was in progress, leading to complaints of 'dressing inappropriately'*" i.e. the Culture (row 2) and Responsibility (row 1) were not addressed until the deployment was underway and there was a cultural and responsibility disconnect between the teachers and the school regardless of the deployment. Project researchers (including undergraduates) developed materials around the themes suggested by the 3rd party and these were delivered by the undergraduates with little scope for teacher influence (i.e. rows 1 & 3) - "*The undergrads designed and presented the materials, we checked it, and the teachers also looked at it*".

The thinkactive project worked alongside an existing 3rd party delivered program on health and fitness, and allowed teachers to set activity challenges to motivate and stimulate activity in the classroom (i.e. more teacher responsibility) - "*The teacher built* *on what was delivered [by the 3rd Party], to extend that across the whole curriculum as the same teacher was with the students all the time*". There was little scope to adapt technology for teachers' purposes (orchestration, measuring progress, assessment etc.) beyond what initial designs (row 3).

The interviewees were also concerned with the legacy of their work, and how it affects the classroom (i.e. 'lasting change' from the Reflexive Modelling column of the matrix) - they aspire to leave their technology deployed beyond the scope of the project, and agreed that the matrix would be useful when monitoring and evaluating this 'from a distance', i.e. by structuring ongoing contact with teachers and providing a framework for co-evaluating the technology. In the microbit project "*It took a while for the schools to realise the kit was theirs to keep, but once they did they were excited about what projects they could do, and the matrix could be used to follow that*", and in think active: "*I'd just like to leave the thing in there for a year and see what the teachers come up with*".

In both cases, the interviewees were introduced to the matrix after their deployments, and thus saw it mainly as a reflective and evaluation tool. However, both identified areas where their planning and monitoring could have been improved if they had prior knowledge of the matrix. Both also intimated that using the matrix as a long-term tool to monitor a "legacy" deployment would be valuable.

### 5.2 Simple Example Use Case

How might the matrix be used to Plan, Monitor, Evaluate and Report on an educational technology deployment in a classroom? Suppose a School, Teacher and Researchers are considering a classroom deployment of a specific Technology.

The first step would be understanding the **coherence** of the parties across the suggested themes (**Responsibility, Culture, Technology for Teacher** and **Resources**), how aligned is the design principles of the Technology with the teaching practices and principles of the School and the Teacher (and indeed how well aligned is the Schools stated principles and the Teachers actual practice?), and how can the School resources be allocated to the deployment? This data can be captured through observation of current practice and interviews between Researchers and the School and Teachers. It is possible at this stage that significant issues with coherence can be identified (possibly leading to changes in design, or in extreme cases abandonment of the deployment).

Depending on the flexibility of the design (are researchers presenting a mature design to test or will there be a co-design process), the level of **Cognitive Participation** for the deployment. It should be established the extent that the parties can influence the initial deployment before the Technology is used. It is at this stage that formal co-design activities could be used, e.g., the SAMR/TPACK models [19,39] can be adapted to form part of the process.

Once the deployment is underway (i.e. the technology is in use), how flexible is the Technology to change through **Collective Action**? (i.e. is change needed, can it be made in timely fashion and will it undermine the agreed design principles of the technology?) Which parties can make these changes and will it impact responsibilities, culture, technology for the teacher and resources?

When evaluating the deployment, what has changed about the classroom practices of the teacher, policies of the school and research outcomes for the research team? Often only the latter is considered, but to be a successful deployment there needs to be some ongoing impact on in-classroom activity (which can be independent of the technology) if changes a teachers NDS have occurred). This **Reflexive Modelling** process allows all parties to assess the impact of the deployment and as the study has followed the structure of the matrix, it provides a structured format for reporting.

## 6    Discussion and Conclusion

Throughout this paper, we have discussed the development of the Classroom Technology Deployment Matrix, how it is based on the combination of teacher's Normal Desired State and Normalization Process Theory, utilising the analysis of two studies describing deployments of the same technology in the classroom. In this section, we discuss how we envision the CTDM will be used. Although not strictly temporal – each part of the model can be used throughout a deployment multiple times, we foresee that the columns of the matrix (i.e. the NPT headings) will be particularly useful in key phases of a deployment.

The first column, Coherence, largely deals with the state of the stakeholders before a deployment, and as such can be used to head-off potential challenges, or even to judge whether a deployment is worthwhile at all – in essence a feasibility study. If there is too much divergence between the actors' initial standpoints, then there is little hope that the deployment would be successful. Indeed, this can uncover hidden pitfalls in a deployment at an early enough stage to counteract them. For example, in our first study, there was an underlying disparity between the teachers' philosophies and the schools stated policies, even before the researchers views are taken into account. Our study fell into the trap of taking the school policy at face value - only discovering this issue part way through the deployment.

The second column, Cognitive Participation, deals with the flexibility of the initial *design*, and the flexibility of the actor's learning philosophies. This can be thought of as a negotiation phase. Stakeholders can decide how far they can move from their initial positions (as explored in column 1), and with what incentive. It is important that all stakeholders can have appropriate input at this stage, as well as understanding their own limits of flexibility. For example in our first study, we presented a 'complete' design (tested in the lab) to the teachers who had very little input. This led to a lack of ownership of the intervention. In the second study, the design was not presented as finished and the teacher could make suggestions – indeed it lead to part of the task being removed so the teacher could create assessable individual tasks, thus giving the teacher some control and ownership of the intervention.

The third column, collective action, covers what can be accomplished during an ongoing deployment, i.e. focusing on the *implementation* rather than the design. During this phase, modifications can be made, provided the mechanisms exist to do so. In our studies, this was manifest through the teacher and researcher plans for each lesson. In the first, the teachers had high-level plans (e.g. topics to cover) but left much to the technology, with little communication and collaboration. In the second, the teacher (with some input from the research team) devised plans in more detail to cover key learning outcomes.

The final column, Reflexive Modelling, deals with evaluation and reflection of the deployment. Importantly, what, if anything, has changed? Aside from learning outcomes, has the school or teacher changed their outlook regarding technology? Have the researchers learned about the realities of classroom teaching? In our second study, the teacher was enthusiastic about technology and keen to think about future incorporation, while in the first study the teachers were more uneasy about what they could gain from the technology. Put simply, we had not integrated with their NDS.

In conclusion, the proposed CTDM is a high-level tool for Planning, Monitoring, Evaluating and Reporting on classroom technology deployments. It is flexible enough to allow for parts of the matrix to be addressed with existing well established models, such as SAMR and TPACK. It is designed to bring together the different (though hopefully cooperative) agendas of the stakeholders in a deployment. Specifically, it incorporates the teacher's Normal Desired State and a more general technology adaptation model, Normalization Process Theory, into a tool targeted specifically at this problem area. The tool can be tested and improved in future deployments, but this work forms the basis of a valuable and useful tool for learning technology researchers, teachers, schools and ultimately students.

As a proposed model, there is scope for refinement and improvement, and a key item for future work would be to analyse the long-term impact of using the matrix to plan, monitor, evaluate and report on an educational technology deployment.

**Conflict of Interest Statement: No known Conflict of Interest for this submission**

**Selection and Participation of Children: This work is based on previously published studies and no new data was acquired. Data is available from institution on request.**

**Ethics: The studies included in this work are previously published and obtained full ethics. No further data was acquired.**

**Funding: No additional funding was obtained for this study.**